\documentstyle[prc,preprint,tighten,aps]{revtex}
\begin{document}
\draft
\preprint{\vbox{\noindent 
 \hfill LA-UR-96-2522 \\
 \null\hfill nucl-th/9608007 }}
\title{$\bbox{S}$-matrix and $\bbox{R}$-matrix
determination of the low-energy $\bbox{^5}$He and
$\bbox{^5}$Li resonance parameters} 
\author{Attila Cs\'ot\'o$^{1,2}$ and G.\ M.\ Hale$^2$} 
\address{$^1$National Superconducting Cyclotron Laboratory,
Michigan  State University, East Lansing, Michigan 48824 \\
$^2$Theoretical Division, Los Alamos National Laboratory, Los
Alamos,  New Mexico 87545}
\date{August 5, 1996}

\maketitle

\begin{abstract}
\noindent
We study the low-energy $3/2^-$ and $1/2^-$ states of $^5$He
and $^5$Li in a microscopic cluster model. The scattering
phase shifts of Bond ($\alpha +n$) and of Schwandt ($\alpha
+p$), respectively, are reproduced well. We determine the
resonance parameters by localizing the poles of the
analytically continued $S$-matrix at complex energies. Our
results differ from conventional $R$-matrix resonance
parameters, which were extracted from experimental data
using the definition of a resonance based on the positions 
and widths of reaction cross section peaks. However, they 
nicely agree with the results of an extended $R$-matrix 
method which works at complex energies.
\end{abstract}

\narrowtext
 
Wigner's $R$-matrix is one of the most powerful tools in
nuclear physics. It is routinely used to analyze experimental
data. Just to mention one example, the analysis of the
$^{12}$C($\alpha,\gamma)$$^{16}$O process, which is a key
reaction in nuclear astrophysics \cite{Fowler}, is heavily
based on $R$-matrix fits \cite{C12ag}. The beauty of Wigner's
method is that all scattering quantities are parametrized in
terms of real, energy-independent quantities.  It is true that
the values of some of the parameters of the theory, namely the
reduced-width amplitudes, $\gamma_{\lambda c}$, and energy
eigenvalues, $E_\lambda$, are somewhat arbitrary due to their
dependence on the boundary-condition numbers, $B_c$, and
channel radii, $a_c$.  However, any sufficiently robust
description ({\em i.e.} one including enough levels) of the
scattering data in a given energy region will give stable
(independent of $a_c, B_c$) resonance parameters when they are
based on the actual complex-momentum poles and residues of the
$S$-matrix, as calculated from continuing the $R$-matrix into
the complex energy plane.

We will call this method \cite{dtan} based on the actual
$S$-matrix pole structure in the complex plane the ``extended"
$R$-matrix prescription for defining resonance parameters, in
order to distinguish it from the usual resonance-parameter
prescriptions that are defined entirely on the real energy
axis.  The real-energy parameters are easier to obtain,
especially when they are extracted directly from the
measurements, which of course exist only on the real energy
axis of the physical sheet.  However, it is sometimes
difficult to interpret experimental results based solely on
methods that work at real energies.  We mention here two
examples: the large cross section of the $t(d,n)\alpha$
reaction and the problem of the soft dipole resonance in
neutron halo nuclei.  Only analyses at complex energies were
able to reveal that the large reaction cross section is caused
by a shadow pole of the scattering matrix in the former case
\cite{dtan,He5}, and that the soft dipole resonance does not exist
in $^6$He in the latter one \cite{soft}. 
 
It is an intriguing question whether or not the results of
methods that are not confined to real energies agree with
those of conventional analyses for the relatively broad
resonances encountered in light systems. As an exploratory
investigation, in this paper we study the low-energy $3/2^-$
and $1/2^-$ states of the $^5$He and $^5$Li nuclei. The
parameters of these states have been determined from
conventional $R$-matrix analyses of certain experimental data.
In those works the definition of the resonance position 
and width was based on the positions and widths of peaks 
in the reaction cross sections. 

For narrow isolated single channel resonances the two definitions 
of resonance parameters ($S$-matrix pole and cross section peak, 
respectively) are consistent with each other, and give the same 
results. However, for 
broad resonances the results coming from the two definitions 
may be different, because only the scattering theoretical 
quantities defined at complex energies (e.g., the $S$-matrix, 
the Fredholm determinant, and the Jost function on the 
multisheeted Riemann energy surface) contain the correct 
dynamical information. Even in the 
case of a narrow multichannel resonance the results 
coming from the two definitions can disagree, 
and studying the complex energy scattering quantities can 
give a much deeper insight into the dynamics of the problem 
\cite{dtan,He5}.  

The extraction of the $\alpha +N$ resonance parameters from
observables which are continued to complex energies was first
suggested by Ahmed and Shanley \cite{Ahmed}. They pointed out
that the determination of the resonance parameters from real
energy observables is difficult because, $e.g.$, the
$1/2^-$ phase shift does not even pass through 90 degrees.
Here we determine these parameters from the analytic
continuation of the $\alpha+N$ scattering matrix to complex
energies in a microscopic model. We also extract the $^5$He
and $^5$Li resonance parameters from the extended $R$-matrix
method.

Our model is a microscopic $\alpha +N$ Resonating Group
Method (RGM) approach to the five-nucleon system. The trial
function of the five-body problem reads
\begin{equation}
\Psi=\sum_{i=1}^{N_\alpha}
{\cal A}\left \{\left [ \left [(\Phi ^{\alpha_i}\Phi^N)
\right ]_S
\chi_L^{\alpha_iN}(\bbox{\rho}_{\alpha N})\right ]_{JM} \right \},
\label{wf}
\end{equation}
where ${\cal A}$ is the intercluster antisymmetrizer, the
$\bbox{\rho}_{\alpha N}$ vector is the intercluster Jacobi
coordinate, $L$ and $S$ is the total angular momentum and
spin, respectively, and [...] denotes angular momentum
coupling. While $\Phi^N$ ($N=n$ or $p$) is a neutron or proton
spin-isospin eigenstate, the antisymmetrized ground state
($i=1$) and monopole excited states ($i>1$) of the $\alpha$
particle are represented by the wave functions
\begin{equation}
\Phi ^{\alpha_i}=
\sum_{j=1}^{N_\alpha}A_{ij}\phi ^\alpha _{\beta _j},\ \
i=1,2,...,N_\alpha.
\label{alpha}
\end{equation}
Here $\phi ^\alpha _{\beta _j}$ is a translationally
invariant shell-model wave function of the $\alpha$ particle
with size parameter $\beta _j$ and the $A_{ij}$ parameters are
to be determined by minimizing the energy of the $\alpha$
particle \cite{Tang}. Putting (\ref{wf}) into the five-nucleon
Schr\"odinger equation which contains a two-nucleon strong and
Coulomb interaction, we arrive at an equation for the
intercluster relative motion functions $\chi$. This equation
is solved by utilizing a Kohn-Hulth\'en  variational method
for the $S$-matrix, which uses square integrable basis 
functions matched with the correct scattering asymptotics
\cite{Kamimura}. 

The input data for the $R$-matrix studies are the cross
sections and polarizations for all possible  reactions
involving $^5$He and $^5$Li. In the present RGM approach we
concentrate on reproducing the $\alpha+N$ scattering phase
shifts  because they are most closely related to the usual
definition of the resonance  position and width. So, if our
model reproduces the phase shifts, then our  resonance
parameters are hopefully close to the ones that characterize
the reactions. Moreover, in order to analyze cross sections,
we should build reaction mechanisms into the model, which
would make this model rather phenomenological and ambiguous.

We use the Minnesota effective $N$--$N$ interaction
\cite{MN}. The same model (for the $^5$He and $^5$Li
subsystems) and interaction were used in \cite{He6,beta,soft}
to successfully describe the structure and beta delayed
deuteron emission of $^6$He, and the three-body resonances of
the $A=6$ nuclei. 

In Fig.\ 1 we show our $S$- and $P$-wave $\alpha +N$ phase
shifts, together with the experimental data of Bond
\cite{Bond} for $\alpha +n$ and Schwandt \cite{Schwandt} for
$\alpha +p$. A rather good agreement is observed, especially
in the resonance region. We do not show the higher partial
waves because they do not influence our results, and they are
practically zero at low energies in agreement with the
experiments.

The experimental parameters of the low-lying $^5$He and
$^5$Li states are listed in Table I. We compare our results
with those of Barker \cite{Barker} and with the compilation
\cite{Ajzenberg}. The results of Ref.\ \cite{Ahmed} are also
shown.

The $3/2^-$ and $1/2^-$ resonance parameters are
determined by analytically continuing the $S$-matrix to
complex energies \cite{He5}. In practice this is done by
solving the Schr\"odinger equation for the $\alpha - N$
relative motion at complex energies with the the following
boundary condition for $\rho_{\alpha N}\rightarrow \infty$ 
\begin{equation}
\chi_L^{\alpha_iN}(\varepsilon_i,\rho_{\alpha N}) 
\rightarrow H_L^-(k_i\rho_{\alpha N})-\tilde
S_L(\varepsilon_i) H_L^+(k_i\rho_{\alpha N}).
\end{equation}
Here $\varepsilon_i$ and $k_i$ are the {\em complex} energies
and wave numbers of the relative motions, and $H^-$ and 
$H^+$ are the incoming and outgoing Coulomb functions, 
respectively. The function $\tilde S$ has no physical 
meaning, except if it is singular at the energy 
$\varepsilon$. Then $\tilde S$
coincides with the physical $S$-matrix describing a purely
outgoing solution, that is a resonance. So we search for the
poles of $\tilde S$ at complex energies. All quantities are
defined on the multisheeted Riemann energy surface, and are
analytic almost everywhere. The complex Coulomb functions were
calculated by using \cite{coulcc}.

The positions ($E_r$) and widths ($\Gamma$) of the resonances
are extracted from the $\varepsilon = E_r-i\Gamma/2$ complex
pole positions of the $S$-matrix. We compare these
parameters with those coming from the cross section peak
definition of a resonance, used in conventional $R$-matrix
approaches. We can see in Table I, that
our parameters are rather different from the $R$-matrix
results of \cite{Barker}. It is especially intriguing that the
splitting of the $3/2^-$ and $1/2^-$ states is much
smaller than in \cite{Barker}, not to mention
\cite{Ajzenberg}. We explored the dependence of the phase
shifts on the resonance parameters by slightly changing the
$N$--$N$ interaction. In Fig.\ 1(a) the dashed lines show the
change of the resonant phase shifts when the resonance
parameters of Table I are changed by 10\%. We also checked the
effect of including the $d+t$ channel in Eq.\ (\ref{wf}), and
found that the resonance parameters are little changed,
provided the phase shifts have the same quality as in Fig.\
1(a).

We also used the extended $R$-matrix method \cite{dtan} to
extract resonance parameters for the $A=5$ ground- and first
excited-states from multichannel $R$-matrix analyses of
reactions in the $^5$He and $^5$Li systems.  The $A=5$
analyses included the two-body channels $N+\alpha$ and $d+t$ or
$d+^3$He, along with pseudo two-body configurations to
represent the breakup channels $n+p+t$ or $n+p+^3$He. 
Included in the $n-\alpha$ data are the differential elastic
scattering cross sections of Morgan \cite{Morg}, Hoop
\cite{Hoop}, Niiler \cite{Niil}, and Shamu \cite{Sham};
polarization and analyzing-power measurements by Sawers
\cite{Saw}, Broste \cite{Bros}, May \cite{May}, and Perkins
\cite{Perk}; and neutron total cross sections measured by
Haesner \cite{KFK}.  The $p-\alpha$ data include the
differential elastic scattering cross sections of Freier
\cite{Frei}, Jarmie \cite{Jarm}, Garreta \cite{Garr}, and
Plattner \cite{Plat}; polarization and analyzing-power
measurements by Schwandt \cite{Schwandt}, Plattner
\cite{Plat}, and Hardekopf \cite {Hard}; and
polarization-transfer measurements by Keaton \cite{Keat}. 

The results are given in Table I. It is remarkable that most
of the resonance  parameters nicely agree with the RGM
results, but differ from the conventional $R$-matrix
results. Perhaps the only exception is the $1/2^-$ state of
$^5$Li,  where the agreement between the RGM and extended
$R$-matrix results is not  so good. The small differences
between the extended $R$-matrix and RGM results  in the case
of $^5$He probably come from the fact that our $R$-matrix
gives  phase shifts that are slightly different from
\cite{Schwandt} at higher  energies. Our RGM results are also
in good agreement with \cite{Ahmed},  where a higher order
scattering amplitude expansion was used.  

In addition to the dependence of the resonance parameters of
Ref. \cite{Barker} on the mechanism by which $^5$He and
$^5$Li are formed, there is also a marked dependence on
channels radius.  Barker has argued that this dependence can
be used to determine a ``best" value of channel radius, which
for $n+\alpha$ is taken to be 5.5 fm.  It is therefore quite
interesting that when the $n-\alpha$ $R$-function parameters
from Table 8 of Ref. \cite{Barker} are used in the $S$-matrix
pole prescription \cite{dtan}, the resulting resonance
parameters, $E_{3/2^-}=0.77$ MeV,
$\Gamma_{3/2^-}=0.65$ MeV; $E_{1/2^-}=2.10$ MeV,
$\Gamma_{1/2^-}=5.37$ MeV, are in good agreement with
the RGM values and with the extended $R$-matrix values,
defined for $a_{n-\alpha}=3.0$ fm.  

The simplest way to extract resonance parameters is to fit
the cross section or phase shift data with Breit-Wigner forms.
This is, however, an ambiguous procedure for broad resonances
where the phase shift is not ``ideal'' ({\em i.e.}, not going
from 0 to 180 degrees within a short energy interval). 
For an ``ideal'', isolated, narrow resonance the phase shift, 
given by scattering theory, behaves like 
$\tan \delta (E)=0.5\Gamma/(E_r-E)$, which implies that 
$d\delta/d E$ has a maximum at $E=E_r$, and $\Gamma=2/(d 
\delta/d E)_{E_r}$. This prescription to extract $E_r$ and 
$\Gamma$ is also used for broader resonances \cite{Joe}.
In Table II we show the $3/2^-$ and $1/2^-$ parameters coming
from this definition applied to the experimental and model 
phase shifts. One can see that this simple procedure provides
resonance parameters which are close to the RGM results,
except that it systematically overestimates the widths. 

In a very recent paper \cite{Efros} the authors extracted the
resonance parameters of the $^5$He and $^5$Li ground states by
determining the pole positions of the $S$-matrix corresponding
to $^3$H($d,\gamma$)$^5$He and $^3$He($d,\gamma$)$^5$Li
measurements \cite{Balbes}. Their results, 
$E_{3/2^-}$($^5$He$)=0.8\pm 0.02$ MeV,
$\Gamma_{3/2^-}$($^5$He$)=0.65\pm 0.02$ MeV, and
$E_{3/2^-}$($^5$Li$)=1.72\pm 0.03$ MeV,
$\Gamma_{3/2^-}$($^5$Li$)=1.28\pm 0.03$ MeV are in good
agreement with our values in Table I.

In summary, we have determined the parameters of the
low-energy $^5$He and $^5$Li resonances from the complex pole
positions of the $\alpha +N$ scattering matrix in a
microscopic cluster model. Our results are different 
from the results coming from a conventional $R$-matrix 
method, which define the resonance
parameters based on the real energy properties of cross
section peaks. However, they are in good agreement with the 
results of an extended $R$-matrix method that works in the 
complex energy plane.  We emphasize that the extended
$R$-matrix method involves no difference in the way that
$R$-matrix parameters are extracted from experimental data,
but only in the way that they are subsequently used to define
resonance parameters.  However, it is quite clear that the
dependence on the channel radius of the usual real-energy
prescription for defining resonance parameters from
$R$-matrix parameters is characteristic of that 
prescription.   

The $S$-matrix pole prescription gives consistent
resonance parameters for the ground state and first excited
state of the $A=5$ nuclei, which are approximately independent
of the method used to describe the nuclear dynamics or the
reaction in which the resonance is observed, and in the case of
the $R$-matrix parametrizations, also independent of the
channel radii and boundary conditions.  We expect this would
also be the case for relatively broad levels in other light
systems, where different resonance-parameter prescriptions can
lead to quite different results \cite{A=4}, and so we
recommend using the complex $S$-pole prescription to specify
resonance parameters in all cases.

\mbox{\ }

We thank Ian Thompson for some useful information on his
complex Coulomb code, and Fred Barker and Jean Humblet for 
enlightening discussions. The work of A.\ C.\ was supported by
Wolfgang Bauer's Presidential Faculty Grant (PHY92-53505)
and by OTKA Grant No.\ T019834. This work was also supported
by NSF Grant No.\ PHY94-03666 (MSU) and by the US Department
of Energy (Los Alamos).

\mediumtext
\begin{figure}
\caption{Phase shifts for $\alpha +n$ (a) and $\alpha +p$ (b)
scattering in our RGM model. The dashed lines are results from
calculations where the resonance parameters differ from those
in Table I by 10\%. Experimental data points are taken from
\protect\cite{Bond} (a) and \protect\cite{Schwandt} (b),
respectively. 20 degrees are added to the $S_{1/2}$ phase
shifts, for clarity.}
\label{fig1}
\end{figure}

\widetext
\begin{table}
\squeezetable
\caption{Parameters of the low-energy $^5$He and $^5$Li
resonances in the center-of-mass frame, coming from the
$S$-matrix pole and cross section peak definitions,
respectively. $E_r$ is the resonance
position relative to the $\alpha +N$ threshold, and $\Gamma$
is the full width at half maximum. All numbers are in MeV.} 
\begin{tabular}{lr@{}lr@{}lr@{}lr@{}lr@{}lr@{}lr@{}lr@{}l} 
& \multicolumn{8}{c}{$^5$He} & \multicolumn{8}{c}{$^5$Li} \\
\cline{2-9}
\cline{10-17}
Method
& \multicolumn{2}{c}{$E_r(3/2^-)$}
& \multicolumn{2}{c}{$\Gamma (3/2^-)$}
& \multicolumn{2}{c}{$E_r(1/2^-)$}
& \multicolumn{2}{c}{$\Gamma (1/2^-)$}
& \multicolumn{2}{c}{$E_r(3/2^-)$}
& \multicolumn{2}{c}{$\Gamma (3/2^-)$}
& \multicolumn{2}{c}{$E_r(1/2^-)$}
& \multicolumn{2}{c}{$\Gamma (1/2^-)$}
\\
\tableline
Compilation \protect\cite{Ajzenberg} &
0.&89$\pm$0.05 & 0.&60$\pm$0.02 & \ \ 4.&89$\pm$1 &
\multicolumn{2}{c}{4$\pm$1} &
1.&96$\pm$0.05& \multicolumn{2}{c}{$\approx$ 1.5}&
\multicolumn{2}{c}{7--12}&
\multicolumn{2}{c}{5$\pm$2}\\
$R$-matrix, stripping \protect\cite{Barker} & 
0.&838$\pm$0.018 & 0.&645$\pm$0.046 & 2.&778$\pm$0.46 &
3.&6$\pm$1.2 & 1.&76$\pm$0.06 & 1.&18$\pm$0.13 &
3.&63$\pm$0.56 & 4.&1$\pm$2.5 \\
$R$-matrix, pickup \protect\cite{Barker} & 0.&869$\pm$0.003 &
0.&723$\pm$0.019 & 3.&449$\pm$0.4 & 5.&3$\pm$2.3 &
1.&86$\pm$0.01 & 1.&44$\pm$0.08 & 4.&54$\pm$0.5 & 6.&1$\pm$2.8
\\ Scattering ampl.\ \protect\cite{Ahmed} &  0.&778 & 0.&639 &
1.&999 & 4.&534 &1.&637 & 1.&292 & 2.&858 & 6.&082 \\
$S$-matrix, RGM& 0.&76 & 0.&63 & 1.&89 & 5.&20 & 1.&67 & 1.&33
&2.&70 & 6.&25 \\
Extended $R$-matrix & 0.&80 & 0.&65 & 2.&07 & 5.&57 & 1.&69 &
1.&23 &3.&18 & 6.&60\\
\end{tabular}
\end{table}

\mediumtext
\begin{table}
\caption{Parameters of the low-energy $^5$He and $^5$Li
resonances in the center-of-mass frame, determined by 
assuming that $d \delta/d E$ has a maximum at $E=E_r$, and 
$\Gamma=2/(d \delta/d E)_{E_r}$. $E_r$ is the resonance position
relative to the $\alpha +N$ threshold, and $\Gamma$ is the
full width at half maximum. All numbers are in MeV.}
\begin{tabular}{lr@{}lr@{}lr@{}lr@{}lr@{}lr@{}lr@{}lr@{}l} 
& \multicolumn{8}{c}{$^5$He} & \multicolumn{8}{c}{$^5$Li} \\
\cline{2-9}
\cline{10-17}
Phase shifts
& \multicolumn{2}{c}{$E_r(3/2^-)$}
& \multicolumn{2}{c}{$\Gamma (3/2^-)$}
& \multicolumn{2}{c}{$E_r(1/2^-)$}
& \multicolumn{2}{c}{$\Gamma (1/2^-)$}
& \multicolumn{2}{c}{$E_r(3/2^-)$}
& \multicolumn{2}{c}{$\Gamma (3/2^-)$}
& \multicolumn{2}{c}{$E_r(1/2^-)$}
& \multicolumn{2}{c}{$\Gamma (1/2^-)$}
\\
\tableline
Exp.\ data \protect\cite{Bond,Schwandt}& 0.&77 & 0.&69 & 2.&13
& 7.&26 & 1.&53 & 1.&42 &2.&77 & 8.&89 \\
RGM& 0.&76 & 0.&68 & 2.&07 & 7.&18 & 1.&67 & 1.&46 &2.&92 &
8.&88 \\
$R$-matrix & 0.&75 & 0.&85 & 2.&21 & 7.&98 & 1.&67 &
1.&37 &3.&35 & 9.&40 \\ 
\end{tabular}
\end{table}

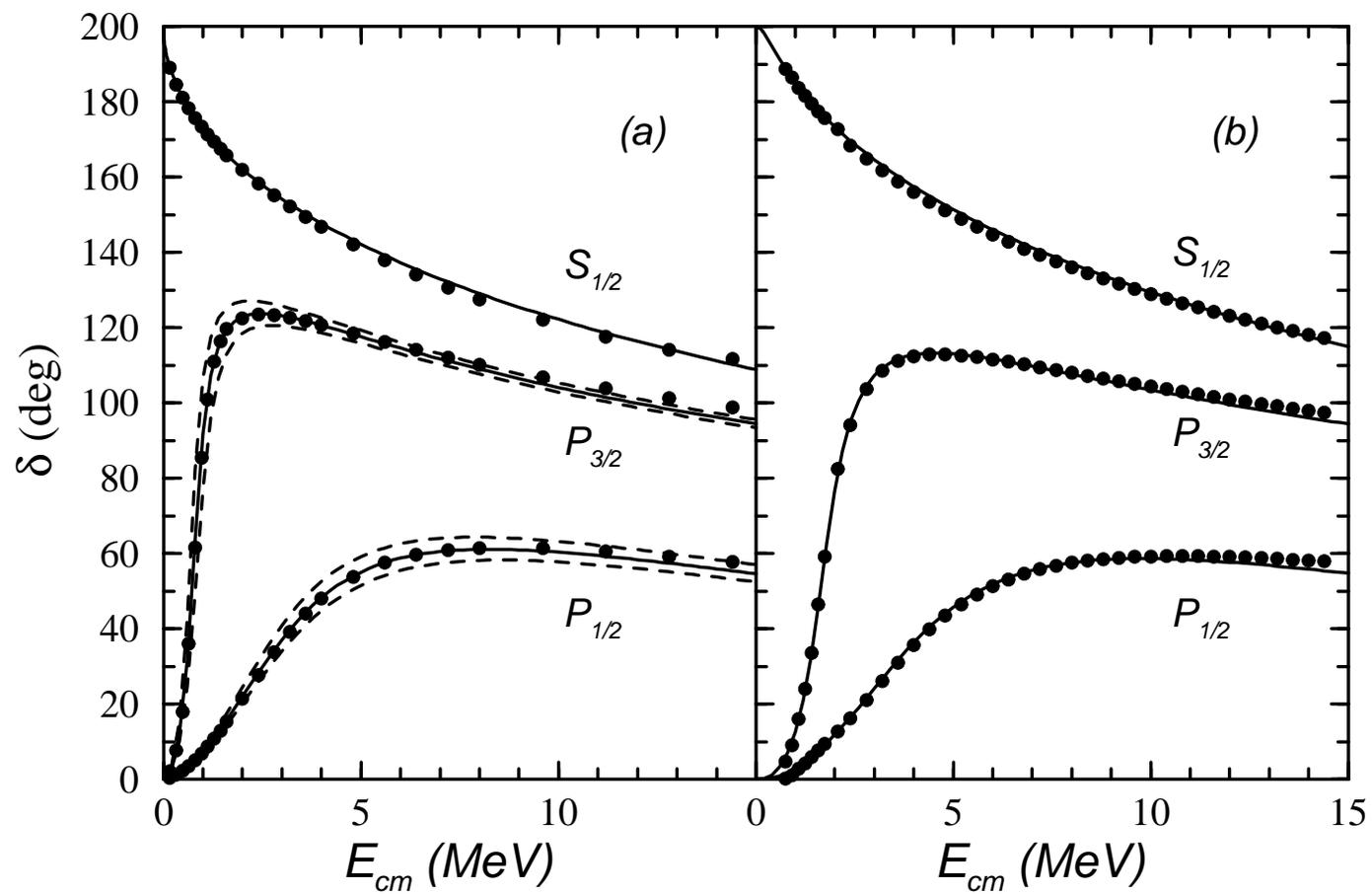
\end{document}